# A U-shaped bilayer graphene channel transistor with a very high $I_{on}/I_{off}$ ratio

Z. Moktadir, S.A. Boden, A. Ghiass, H. Rutt. and H. Mizuta

A novel graphene transistor architecture is reported. The transistor has a U-shape geometry and was fabricated using a gallium focused ion beam (FIB). The channel conductance was tuned with a back gate. The $I_{on}/I_{off}$ ratio exceeded $10^5$.

Introduction: Graphene has ignited tremendous interest since its discovery in 2004[1]. This material is made of a single layer of carbon atoms assembled in a hexagonal structure, providing extraordinary electrical and mechanical properties. The observation of field effect in graphene has particularly attracted the attention of the electron devices community resulting in a large volume of work on graphene field effect transistors (GFET). Indeed graphene is one of the possible materials for the neo silicon era, according to the 2009 International Technology Roadmap for Semiconductor . One outstanding property of graphene is its high carrier mobility at room temperature. In exfoliated supported graphene, mobility as high as 70000 $cm^2$ $V^{-1}$ $s^{-1}$ has been reported [2]. Suspended graphene shows even higher values, reaching $10^6$ $cm^2$ $V^{-1}$ $s^{-1}$ as the interaction with the substrate is suppressed [3]. Graphene grown on metals and transferred to a substrate shows lower values . However, large area monolayer or bilayer graphene suffers from the absence of a band gap, which results in poor switching capability of GFETs. Despite the intense work to improve the $I_{on/off}$ ratio, only values less than 100 were achieved. There are two ways used to open a band gap in graphene. The first is to cut it into narrow ribbons (graphene nanoribbons, GNRs). If the width of the ribbon is small enough (few nanometres), carriers are confined in a 1 dimensional quantum wire, resulting in a non-zero band gap. The band gap value is inversely proportional to the width of the

GNR. Band gaps larger than 0.5 eV (needed for device operation at room temperature) require the width of the GNR to be less than 5 nm [4]. Unfortunately existing lithographic techniques cannot provide such accuracy. The second way is to apply a strong electric field perpendicular to the bilayer graphene plane, inducing a strong coupling between the two graphene atomic layers, which results in the detachment of the valence and the conduction band at the K point in the Brillouin zone. The largest band gap values were achieved with dual gated GFETs which are controlled using a top gate and a back gate. Values of $I_{on}/I_{off}$ ratios of around 100 have been demonstrated [5].

We report here a new GFET architecture. The GFET is U-shaped and fabricated using state of the art gallium FIB technology. The obtained $I_{on}/I_{off}$ exceeds $10^5$.

Device fabrication: All fabrication was carried out using a dual focused ion beam/scanning electron microscope system (Zeiss NVision40 FIB/FEGSEM) equipped with a gas injection system (GIS). A bilayer graphene flake on a 300 nm layer of $SiO_2$ on Si (sample purchased from Graphene Industries Ltd.) was imaged (Figure 1a) and areas were identified for the formation of a graphene U-shaped transistor. Electron beam- induced tungsten deposition was then used to deposit thin protective layers across the edges of the graphene flake in areas where contact formation would take place. This process involved using the reduced raster scan to scan the electron beam over these edge areas. The valve to the GIS was then opened to allow tungsten precursor ($W(CO)_6$) to flow into the chamber and deposit on the surface, leading to the controlled deposition of a few nanometres of tungsten across the edges of the graphene flake. Square contact pads, 50 μm x 50 μm, and smaller rectangular strips to connect the pads to the graphene were deposited using gallium ion beam-induced tungsten deposition in the same system. The ion beam current was adjusted in the range 0.08 – 13 nA for maximum deposition rate depending on the areas over which the deposition was occurring. Care was taken not to expose the unprotected parts of the graphene flake to the Gallium ion beam during deposition so as to minimise damage to the graphene. After each deposition, the gas injection system valve was closed and the chamber vacuum was allowed to recover so that any remaining precursor could desorb from the surface before imaging the sample. The Gallium ion beam was then used to mill the contact wires and graphene flake to form and isolate the device. A beam current of 150 pA was used when milling the tungsten and 80 pA when milling the graphene; the Ga+ beam voltage was 30 kV. The samples were annealed at 400 °C for 10 minutes prior to their characterisation.

Results: Figure 1–a shows the unpatterned graphene flake where the dashed outline marks the pristine area where the device is fabricated. Figure 1-b is a low magnification image of the whole device with tungsten contact pads. Only the bottom two contact pads connect to the U-shaped transistor which is the subject of this letter. The upper contact pad, which connects to a second device, can be ignored. The dashed outline marks the same area as in Figure 1a. Figure 1-c and 1-d show higher magnification images of the fabricated GFET. The width of the graphene wire was around 300 nm while the total length was 20 µm. The GFET was controlled by a Si/SiO$_2$ substrate back gate.

Figure 2–a shows the drain current $I_d$ as a function of the back gate voltage, $V_g$, for a drain voltage of 1V. Figure 2-b shows the drain current as a function of the drain voltage $V_d$ for $V_g$ =4 V. The $I_d$-$V_g$ characteristic shows a marked dip around $V_g$=$V_{off}$= 3V, where the GFET channel is switched off. This characteristic is dissimilar to the usual large area graphene $I_d$-$V_g$ characteristic which shows a smooth evolution of the drain current towards the Dirac point. The $I_{on}/I_{off}$ derived here is around 4.8×10$^5$ (taking $I_d$=3.5×10$^{-4}$ A at $V_g$=-5 V, and $I_d$=7.25×10$^{-10}$ A at $V_g$=3.3 V), larger by 3 orders of magnitude than the highest $I_{on}/I_{off}$ obtained in dual gated GFET . The U-shaped GFET shows similar bipolar behaviour observed in usual GFETs, but it also shows

an asymmetry in the two conduction regimes (holes and electrons conduction). Our results can be interpreted by the effect of geometrical corners in the graphene channel. Theoretical calculations predict that the formation of a band gap of an L-shaped channel is associated with quantum-mechanical quasi-bound states at the corners. These quasi bound states are strongly dependent on the atomic configurations on both inner and outer corners [6]. One should not exclude the formation of a mobility gap [7], however, it is unlikely to be the dominant effect as our fabricated channel show very low off-current (~1 nA). The result presented here will open up possibilities for the development of efficient graphene logic circuits. For example, two GFETs with two different $V_{off}$ points can be fabricated side by side to form a complementary GFET (CGFET), with little overlap between the $I_d$-$V_g$ characteristics of the two GFETs, resulting in an efficient inverter.

Conclusion: we have fabricated a graphene field effect transistor having a U-shape geometry using FIB technology. The achieved $I_{on}/I_{off}$ exceeds $10^5$.

**Authors' affiliations:**

University of Southampton, School of Electronics and Computer Science, Highfield, Southampton SO17 1BJ, United Kingdom

Email address: zm@ecs.soton.ac.uk

**Figure captions:**

Fig. 1: Scanning electron micrographs showing: (a) The graphene flake. The dashed outline marks the areas suitable for formation of the devices; (b) The device and the contact pads, with the dashed outline marking the same area as in (a); (c) the graphene U-shaped device with tungsten contact wires and pads, (d) detail of the graphene U-shaped device.

Fig. 2: The $I_d$-$V_g$ characteristic at $V_d$ = 1 V, showing the off-point of the U-shaped transistor (a), and the $I_d$-$V_d$ characteristic at $V_g$ =-4 V (b).

Figure 1

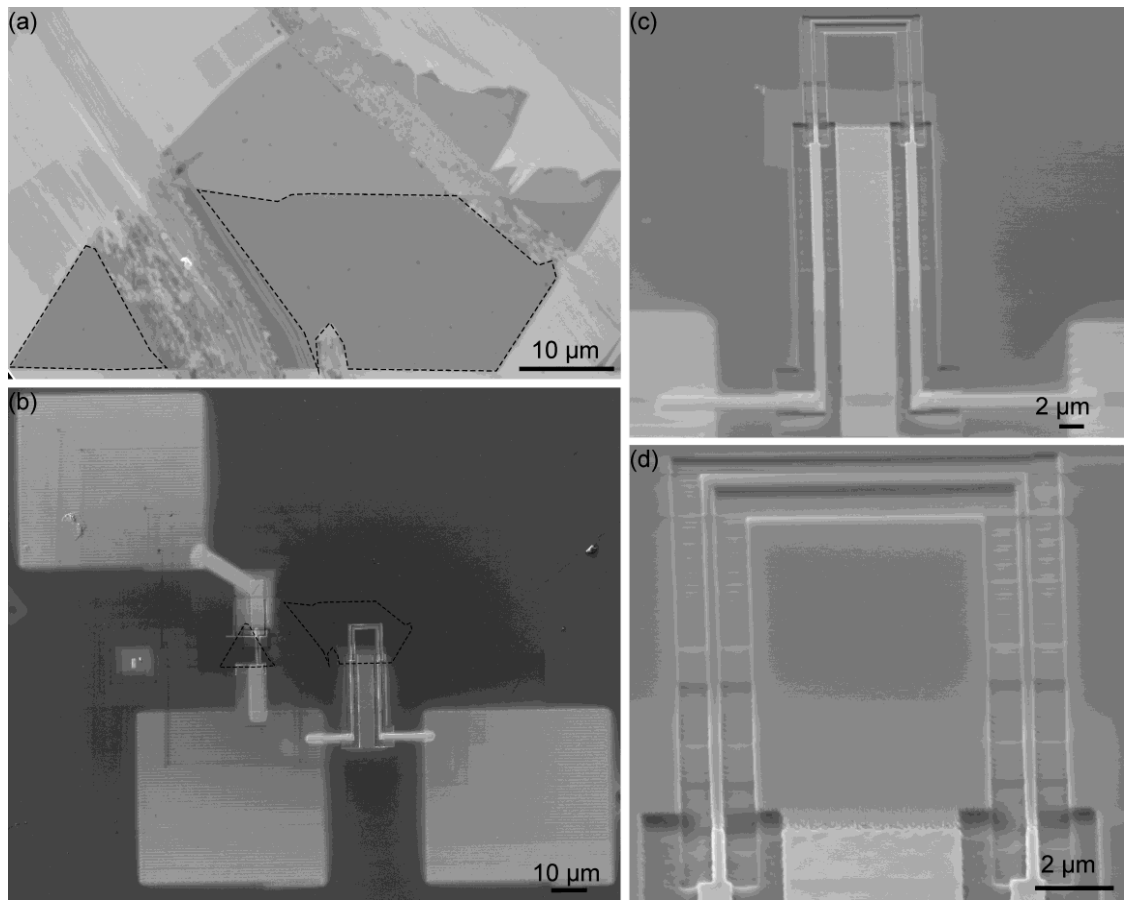

Figure 2

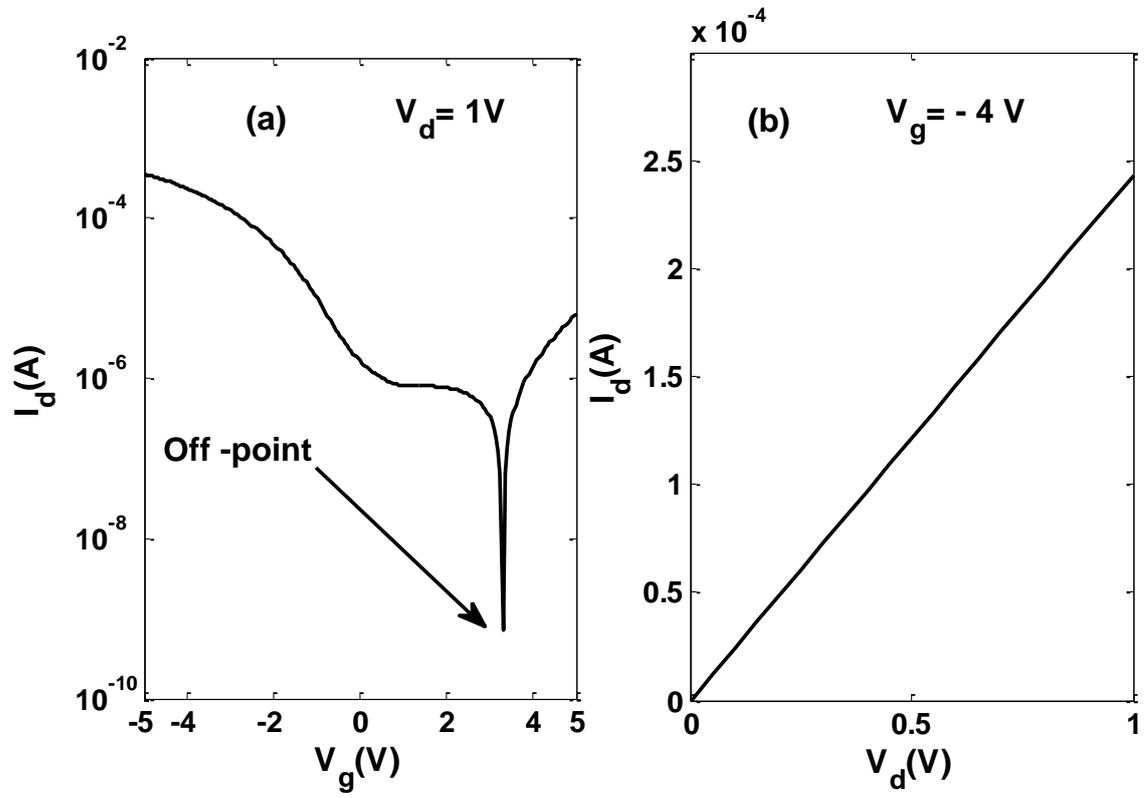